\documentstyle[multicol,prb,aps]{revtex}

\begin{document}

\draft

\title{Oxygen adsorption on the Ru\,(10$\bf\bar{1}$0) surface:
Anomalous coverage dependence}

\author{S. Schwegmann, A. P. Seitsonen, V. De Renzi, H. Dietrich, H.
Bludau, M. Gierer, H. Over\footnote{Corresponding author: e-mail:
over@fhi-berlin.mpg.de; Fax: ++49-30-84135106.}, K. Jacobi, M.
Scheffler, and G. Ertl}
\address{Fritz-Haber-Institut der
Max-Planck-Gesellschaft, Faradayweg 4-6, D-14195 Berlin, Germany}

\maketitle

\vspace{5mm}

\centerline{submitted to Phys. Rev. B (8. Aug. 97)}

\vspace{3mm}

\begin{abstract}
  Oxygen adsorption onto Ru\,(10$\bar{1}$0) results in the formation
of two ordered overlayers, {\it i.\,e.} a c(2$\times$4)-2O and a
(2$\times$1)pg-2O phase, which were analyzed by low-energy electron
diffraction (LEED) and density functional theory (DFT) calculation. In
addition, the vibrational properties of these overlayers were studied
by high-resolution electron loss spectroscopy. In both phases, oxygen
occupies the threefold coordinated hcp site along the densely packed
rows on an otherwise unreconstructed surface, {\it i.\,e.} the O atoms
are attached to two atoms in the first Ru layer Ru(1) and to one Ru
atom in the second layer Ru(2), forming zigzag chains along the
troughs. While in the low-coverage c(2$\times$4)-O phase, the bond
lengths of O to Ru(1) and Ru(2) are 2.08~\AA\ and 2.03~\AA,
respectively, corresponding bond lengths in the high-coverage
(2$\times$1)-2O phase are 2.01~\AA\ and 2.04~\AA (LEED). Although the
adsorption energy decreases by 220~meV with O coverage (DFT
calculations), we observe experimentally a shortening of the Ru(1)-O
bond length with O coverage. This effect could not be reconciled with
the present DFT-GGA calculations. The $\nu$(Ru-O) stretch mode is
found at 67~meV [c(2$\times$4)-2O] and 64~meV [(2$\times$1)pg-2O].
\end{abstract}

\begin{multicols}{2}

\section{Introduction}

  The actual adsorption geometry of atoms and molecules at metal
surfaces, {\it i.\,e.} the adsorption site and the bond lengths of an
adsorbate to its attached substrate atoms, is essentially the result
of a delicate balance between reaching the optimum surface charge
density by the adsorbate, minimizing the Pauli repulsion between
occupied orbitals of the adsorbate and the metal surface, and
optimizing the electrostatic interaction between the adspecies [1]. If
the charge density at the surface is modified, for example, due to
coadsorbed atoms or molecules, the adsorption geometry of the
pre-adsorbed species on the surface can vary widely. Illustrative
examples have only recently been reported in the literature in that
adsorption sites have shown to switch upon coadsorption [2]. Not only
in heterogeneous ({\it i.\,e.} coadsorption) systems but also in
homogeneous systems ({\it i.\,e.} for a single adsorbate) the
adsorption geometry might change upon varying the density of
adparticles since both the effective charge density at the surface and
the adsorbate-adsorbate interaction change appreciably. Notable
variations in the adsorption geometry of such homogeneous systems have
been reported for alkali metal adsorption: For instance, with
increasing coverage the adsorption site shifted either from on-top to
hcp sites, as for the case of Cs on Ru\,(0001) [3a], or the
coordination of adsorption was preserved, but still the adsorption
site changed from fcc to hcp, as encountered with the system
K/Ru\,(0001) [3b]. Similar effects were identified with DFT
calculations for alkali metal adsorption on the (111) and (100)
surfaces of Al [4].

  With oxygen adsorption a change in adsorption site with coverage has
not been identified so far, although for the O/Co\,(10$\bar{1}$0)
system such a change was proposed on the basis of recent scanning
tunneling microscopy (STM) and low-energy electron diffraction (LEED)
investigations [5]. On Co\,(10$\bar{1}$0) oxygen forms a low-coverage
c(2$\times$4)-2O phase, which upon heating transforms irreversibly
into a (2$\times$1)-1O phase, and a high-coverage (2$\times$1)-2O
phase [6]. There is, however, only scant evidence for O atoms changing
their adsorption site from fcc to hcp when going from the
c(2$\times$4)-2O to the (2$\times$1)-1O overlayer; threefold
coordinated fcc and hcp sites are characterized by one and two
substrate atoms in the topmost layer, respectively. While the
chemisorption of oxygen into hcp sites in the (2$\times$1)pg-2O phase
has been recently confirmed with LEED [7], the fcc adsorption site of
c(2$\times$4)-2O still remains speculative, as it is only concluded
from the high intensity of the LEED superstructure spots, which was
interpreted in terms of strong lateral distortions of the substrate
atoms [5]. On the Ru\,(10$\bar{1}$0) surface, which is much easier to
prepare and to clean than Co\,(10$\bar{1}$0), oxygen adsorption also
leads to the formation of c(2$\times$4)-2O and (2$\times$1)pg-2O
overlayers [8]. However, the low O coverage phase on
Ru\,(10$\bar{1}$0) is the thermodynamically stable one, while on
Co\,(10$\bar{1}$0) the c(2$\times$4)-2O phase serves just as a
precursor state for the stable (but heavily reconstructed [9])
(2$\times$1)-1O surface. Neither the clean Ru\,(10$\bar{1}$0) nor the
oxygen-covered surfaces have been subjected to a LEED analysis so far.
The motivation for the project discussed in this paper was to learn
about the interaction of oxygen with the underlying Ru\,(10$\bar{1}$0)
surface and about the modification of the local bonding configuration
upon increasing the O coverage. For this purpose, we applied LEED for
the determination of the atomic geometries, high resolution electron
loss spectroscopy (HREELS) for examining the vibrational properties,
and density functional theory (DFT) calculations to determine the
energetics at this surface.

\section{Experimental and calculational details}

\subsection{HREELS}

  The HREELS measurements were performed in a second UHV apparatus
with a base pressure of 2$\times 10^{-11}$~mbar. The apparatus
consisted of two chambers. The upper chamber contained an argon ion
gun, a quadrupole mass spectrometer, and a LEED optics. The lower
chamber housed a HREELS spectrometer for recording the vibrational
spectra, capable of an energy resolution of $\Delta$E = 1~meV in the
reflected beam, was used at $\Delta$E = 1.9~meV across the sample with
typical count rates of 3$\times 10^5$ counts per second in the elastic
peak. The HREEL spectra were all taken at a 60$^\circ$ angle of
incidence with respect to the surface normal and in specular geometry;
the sample temperature was 293~K. The energy resolution was set to be
1.9~meV. In order to allow for direct comparison of HREELS and LEED
results, the same sample and the same cleaning protocol were used.

\subsection{LEED}

  The LEED experiments were conducted in an ultrahigh-vacuum chamber
(with base pressure 2$\times 10^{-10}$~mbar) equipped with a four-grid
LEED optics, Auger electron spectroscopy (AES) and facilities for
surface cleaning and characterization. The LEED intensity data were
recorded at 110~K using a video-LEED system [10]. The recorded LEED
data were fed into a full-dynamical LEED program developed by Moritz
[11] which is also equipped with a least-squares optimization scheme
[12] in order to perform the simultaneous and automated refinement of
structural (as well as non-structural) parameters. The degree of
agreement between calculated and experimental data was judged by the
reliability factors R$_{\rm P}$ [13] and R$_{\rm de}$ [14] which were
also the quantities to be minimized in the optimization scheme. The
scattering from Ru and O was treated by using up to nine phase shifts
which were corrected for thermal vibrations by employing Debye
temperatures of 420~K for Ru and 450~K for O. These temperatures were
not refined. The phase shifts have already been used in a previous
LEED analysis of the (1$\times$1)-O surface structures of Ru\,(0001)
[15]. The LEED analysis was carried out in two steps. First, an
exhaustive grid search over a wide range in parameter space was
conducted for both O phases with the unrelaxed substrate and the
oxygen-ruthenium interlayer spacing being the only structural
parameter. In the next step, starting from the optimum parameter
values found by the grid searches, automated structure refinements
were carried out. Apart from the first three layer spacings, lateral
and vertical displacements of Ru atoms in the first and second layer
(preserving the corresponding local symmetry of the adsorbate) were
simultaneously and automatically refined.

  In both chambers (HREELS and LEED), the Ru\,(10$\bar{1}$0) sample
was cleaned by argon ion bombardment at 1~keV followed by cycles of
oxygen adsorption and thermal desorption in order to remove surface
carbon. Final traces of oxygen were removed by flashing the surface to
1530~K, resulting in a sharp (1$\times$1) LEED pattern (cf. Fig.~1)
and no impurity losses in HREELS. The phases of c(2$\times$4)-2O and
(2$\times$1)-2O were prepared by exposing the clean Ru\,(10$\bar{1}$0)
at room temperature to 0.7~L and 2.5~L oxygen, respectively, (cf.
Fig.~1). From AES measurements, the ratio of global oxygen coverages
in the two ordered overlayers was 1~:~2. Together with the observation
of a glide plane symmetry in the (2$\times$1) phase, one can safely
assume that both c(2$\times$4) and (2$\times$1) structures contain two
O atoms in the unit cell. At room temperature, the (2$\times$1) LEED
pattern exhibits a glide symmetry plane along the [1$\bar{2}$10]
direction, as inferred from the missing fractional-order spots
(n+1/2,~0), n~= 0,~$\pm$1,~\ldots at normal electron incidence; the
proper nomenclature for this oxygen phase is therefore
(2$\times$1)pg-2O.

  Exposing a (2$\times$1)pg-2O structure to NO$_2$ at 500~K sample
temperature, we tried to prepare an ordered O overlayer structure with
coverage exceeding 1~ML. The same procedure has already been used
successfully for the formation of the (1$\times$1)-O structure on
Ru\,(0001) [15]. On Ru\,(10$\bar{1}$0), however, this procedure leads
only to a streaky (1$\times$2) phase in a wide O coverage range from
1.2~ML to 3\ldots 4~ML as estimated from AES and TDS measurements.
Obviously, a (1$\times$1)-2O is not the most stable configuration at
Ru\,(10$\bar{1}$0) under these experimental conditions, but it may
exist as a metastable phase (cf. the DFT calculations below).

\subsection{DFT calculations}

  The density functional theory (DFT) calculations were performed
using the generalized gradient approximation (GGA) of Perdew et al.\
[16] for the exchange-correlation functional. The action of the core
electrons on the valence electrons was replaced by norm-conserving,
fully relativistic pseudo potentials generated by the scheme of
Troullier and Martins in the fully separable form [17]; the electronic
wave functions were expanded in a plane-wave basis set. The used
cut-off energy for the plane wave expansion of 50~Ry is sufficient to
reliably give the adsorption energies, although the O-O interaction is
not fully converged [18], even with a large pseudo potential core
radius of r$_{\rm c}^{\rm l=0,1}$ = 1.45 bohr; l=1 was used as the
local component. The core radii for the Ru pseudo potential are
r$_{\rm c}^{\rm l=0,2}$ = 2.48 bohr and r$_{\rm c}^{l=1}$ = 2.78 bohr;
l=0 was used as the local component [19]. The k-point sampling of the
surface Brillouin zone was accomplished with an equidistant
8$\times$10 point Monkhorst-Pack grid [20] in the (1$\times$1) unit
cell, giving 20 k-points in the irreducible part of the (1$\times$1)
Brillouin zone; special care was taken to ensure an equivalent
sampling in all (surface) geometries studied. To stabilize the
Brillouin zone integration the occupation numbers were broadened using
a Fermi function with a width of 0.1~eV; the total energies were
extrapolated to the case of no broadening. The surface was modeled
using the supercell approach, using eight layers of Ru\,(10$\bar{1}$0)
and placing the O atoms on one side of this slab. We account for the
difference of the asymptotic electrostatic potential by employing a
surface dipole correction [21]. The calculation scheme allows for
relaxation of the electrons and atoms, where we relaxed the positions
of the O atoms and the atoms in the top two Ru layers, keeping the
lower five Ru layer spacings fixed at the bulk values. A similar
procedure has shown to work reliably for the case of nitrogen adsorbed
on Ru\,(0001) [22].

\section{Results} 

\subsection{LEED results}

  The analyses of the clean Ru\,(10$\bar{1}$0) surface and the
oxygen-induced c(2$\times$4)-2O and (2$\times$1)pg-2O overlayers were
based on experimental data sets containing cumulative energy ranges of
2165~eV, 4525~eV, and 3621~eV, respectively. In comparison to the
oxygen Co\,(10$\bar{1}$0) system, one would not expect to find heavy
reconstructions at the surface (recall that Ru is a much harder
material than Co). Assuming only high-coordination adsorption sites
for oxygen, we are left with eight essentially different models for
the c(2$\times$4)-2O structure as compiled in Fig.~2. The presence of
the glide symmetry plane in the (2$\times$1)pg-2O overlayer imposes
constraints to the structure which further narrows down the number of
possible models depicted in Fig.~3. We should note that the glide
symmetry plane disappeared reversibly upon cooling below 230~K; this
interesting issue will be the subject of a future paper [23]. For this
reason we took the LEED IV data of the (2$\times$1)pg-2O phase at
250~K.

  The clean Ru\,(10$\bar{1}$0) surface was analyzed first. Two
different terminations of the (10$\bar{1}$0) surface are possible,
exhibiting different corrugations of about 0.8~\AA\ (short
termination) and 1.6~\AA\ (long termination), respectively. From a
comparison with the surface geometries of Re(10$\bar{1}$0) [24] and
Co\,(10$\bar{1}$0) [25], we anticipated that only the
short-termination with a small corrugation will be the stable one. The
measurements of LEED IV curves at 110~K turned out to be complicated,
due to small amounts of contaminants arising from the residual gas
adsorption, as the LEED IV curves changed quite substantially after a
few minutes. Yet, using these IV curves for the LEED analysis, we
ended with an optimum structure for which the topmost Ru layer spacing
is almost bulk-like. This finding conflicts with the results of about
10~\% contraction obtained for the topmost Re and Co layer distance on
(10$\bar{1}$0). The best fit was achieved with the expected short
termination giving an overall Pendry R-factor of 0.25. From AES
measurements, which indicated a clean surface, we concluded that very
likely hydrogen, which is inevitably present in the residual gas,
should be responsible for this effect. Therefore, we recorded a
further set of LEED IV data at 430~K, a temperature at which hydrogen
is not stabilized at the surface. With these new LEED data a much
better fit to the experimental data was possible, {\it i.\,e.} R$_{\rm
P}$ = 0.18, and in addition, the structural parameters were now
consistent with corresponding results for Co and Re. The topmost Ru
layer spacing turned out to be contracted by 10$\pm$1.5~\% followed by
a small expansion of 2.5~\% of the second Ru layer spacing in good
agreement with the DFT calculations (cf.\ section 3.3). Structural
investigations of the hydrogen adsorption on Ru\,(10$\bar{1}$0) for
various exposures are underway and will be presented in a forthcoming
paper [26].

  Next, we focus on the atomic geometry of the c(2$\times$4)-2O phase
on Ru\,(10$\bar{1}$0). The various model structures considered in this
LEED analysis are summarized in Fig.~2; all of these models provide at
least one mirror plane across the densely-packed Ru rows in
[1$\bar{2}$10] direction. Only high-symmetry adsorption sites were
tested. The best r-factors reached with these models are listed in
Table~1, from which it becomes clear that the model with oxygen
sitting in so-called hcp sites is preferred. The optimum adsorption
geometry is presented in Fig.~4, and the agreement between
experimental and calculated LEED data can be judged from Fig.~5; the
overall r-factor is R$_{\rm P}$ = 0.26. The chemisorption of oxygen
induces only small lateral (up to 0.05~\AA) and vertical (up to
0.03~\AA) displacements of atoms in the top double layer. The lateral
arrangement of the oxygen atoms provides some clues about the
interaction among the adsorbates. The oxygen atoms form zigzag chains
along the troughs. Zigzag and zagzig chains are separated by an empty
trough establishing the c(2$\times$4) symmetry. The alternation of
zigzag and zagzig chains is necessary to impose c(2$\times$4)
symmetry, otherwise a primitive (2$\times$2) structure would have been
formed. This indicates a long-range interaction between the O chains
even across the densely-packed Ru trenches. The energy of this
interaction is, however, quite small since at temperatures above 550~K
the c(2$\times$4)-O structure disorders as indicated by LEED while the
(2$\times$1)pg persists up to desorption. The appearance of empty
troughs between consecutive zigzag chains, on the other hand, may be
the result of the affinity of oxygen to bind to two Ru atoms in the
topmost Ru layer without sharing these atoms with other O atoms. Last,
the formation of zigzag chains even at low coverages indicates
repulsion between the O atoms sitting on nearest-neighbor sites and
attraction between next-nearest-neighbor sites. This arrangement
maximizes the separation between oxygen atoms within the troughs,
although the overall O-O separation is not maximized; from this point
of view model hcp-c, cf. Fig.~2, would be more favorable. The bond
lengths of oxygen to first-layer Ru atoms and second-layer Ru atoms is
(2.08$\pm$0.06)~\AA\ and (2.03$\pm$0.06)~\AA, respectively. The
contraction (10~\%) of the topmost Ru layer spacing of the clean
Ru\,(10$\bar{1}$0) is partly lifted upon adsorption of oxygen,
resulting in a contraction of about 4~\%.

  These structural characteristics of the c(2$\times$4)-2O phase are
to be compared with the adsorption geometry of the second ordered
oxygen overlayer, {\it i.\,e.} the (2$\times$1)pg-2O. For modeling
this oxygen overlayer, only three different models (Fig.~3) have to be
considered. The corresponding optimum r-factors are compiled in
Table~2. Clearly, also here the hcp site is most favored. The actual
adsorption geometry determined by LEED is presented in Fig.~6, and a
comparison between experimental and calculated LEED data is depicted
in Fig.~7 (the overall r-factor is R$_P$~= 0.25). The presence of the
glide symmetry plane determines the lateral arrangement of the O atoms
to consist again of O zigzag chains along the [1$\bar{2}$10]
direction. The main differences to the c(2$\times$4)-2O configuration
are the absence of empty troughs and the exclusive occurrence of
zigzag chains (and no zagzig chains). Therefore, the transformation of
the c(2$\times$4)-2O into the (2$\times$1)pg-2O phase upon adding
oxygen is accomplished by filling up the empty troughs by O zigzag
chains and shifting every zagzig chain of the c(2$\times$4)-2O along
the [1$\bar{2}$10] direction by one lattice unit. In contrast to the
c(2$\times$4)-2O surface, the high-coverage (2$\times$1)pg-2O phase is
thermally very stable and exists up to desorption. Besides these
general features of the oxygen arrangement, the bond lengths between
oxygen and the first-layer and second-layer Ru atoms amount to
2.01$\pm$0.06~\AA\ and 2.04$\pm$0.06~\AA, respectively. As with the
c(2$\times$4)-2O, oxygen induces only little local reconstructions in
the top Ru double layer. The topmost Ru layer spacing is now slightly
expanded by 4~\%.

\subsection{HREELS Results}

  In order to study the vibrational properties of chemisorbed oxygen
on Ru\,(10$\bar{1}$0), HREEL spectra were recorded in a separate UHV
apparatus. A characteristic set of spectra is shown in Fig.~8. The
main energy loss is found at 64 to 67~meV and is assigned to the Ru-O
stretch mode perpendicular to the surface $\nu$(Ru-O). With
progressing oxygen exposure, this mode shifts from 67~meV down to
64~meV. At an oxygen dose of 10~L, it broadens and obviously contains
several contributions. The mode with a polarization perpendicular to
the surface is expected to be observable with HREELS for a chemisorbed
atom. The observed energy compares well with the value of 64~meV found
for the low-coverage mode ($\theta_{\rm O}$~= 0.25) of atomic oxygen
on Ru\,(0001).

  Between 70 and 100~meV, a broad band of losses is present, which
exhibits peaks at 86 and 95~meV for the 2$\times$1 structure. The loss
at 95~meV occurs already for the smallest dose of 0.1~L. In analogy to
the work of Mitchell and Weinberg [27], we tentatively assign these
losses to subsurface oxygen. Mitchell and Weinberg observed a strong
peak at 80~meV accompanied by a broad band between 80 and 130~meV with
weak peaks at 92 and 102~meV, after dosing the oxygen-covered
Ru\,(0001) surface at $\theta_{\rm O}$~=~0.5 with additional NO$_2$.
They interpreted this as the beginning of RuO$_x$ formation. It should
be noted, however, that a single loss at 81~meV can be prepared
without any broad band at higher energies and that this spectrum is
characteristic for the (1$\times$1)-O overlayer [28]; the atomic
geometry of the latter was recently determined by total energy
calculations and LEED [15]. From these observations we conclude that
on Ru\,(10$\bar{1}$0) - different to Ru\,(0001) - the oxygen atoms
penetrate into the subsurface region right away from the beginning of
oxygen exposure.

  Besides the main $\nu$(Ru-O) stretch mode and the features at higher
energies, peaks at 13--15, 23, 44 and 54~meV are also observed. These
modes are here not analyzed in detail, but it is clear that they
belong to phonons of the oxygen-modified Ru\,(10$\bar{1}$0) surface
and the translational modes of oxygen. For Ru\,(0001) it was recently
discussed [28] that the change in symmetry with oxygen adsorption
phonon bands can be folded back to the $\Gamma$ point and can become
visible in HREELS. The weak peak around 250~meV is assigned to the
$\nu$(C-O) stretch mode from background CO.

\subsection{DFT Calculations}

  The lattice parameters were calculated using the (10$\bar{1}$0)
plane as the base of the unit cell and filling the cell with four
atomic layers according to the stacking sequence. The results
obtained, {\it a} = 2.78~\AA\ and {\it c/a}~= 1.58, are very close to
our values for the Ru\,(0001) surface, thus confirming a good k-point
sampling. The overestimation of {\it a} by about 2~-~3~\%, compared to
the experiment, is found in DFT-GGA calculations for later 4d
transition metals.

  As, to the best of our knowledge, this is the first DFT calculation
for the clean Ru\,(10$\bar{1}$0) surface, we shall elaborate this
issue here to some extent. The clean Ru surface was modeled using an
eight-layer slab, and the two first substrate layers were relaxed. The
hcp\,(10$\bar{1}$0) surface is a more open surface than the hexagonal,
close-packed \,(0001) surface, and there are two possible
terminations. As expected (due to higher coordination of surface atoms
and smaller surface corrugation), the short termination [outermost
layer distance d~= 1/(2$\sqrt{3}$)$\times$c] is energetically favored
by 41~meV/\AA$^2$ over the long-terminated surface [d~=
1/$\sqrt{3}\times$a]. The surface energy of the short termination is
calculated as 176~meV/\AA$^2$ which is $\approx$\,15~\% larger than
our DFT-GGA surface energy for the Ru\,(0001) surface
(154~meV/\AA$^2$). Note that DFT-GGA yields lower surface energies
than DFT-LDA and it appears that LDA is in better agreement with
experiments [29]. The relaxations of the first and second layer turned
out to be d$_{12}$/d$_0$~=~$-$13.7~\% and d$_{23}$/d$_0$~=~$-$0.7~\%,
thus, a slightly larger inward relaxation is obtained than from the
LEED intensity data (d$_{12}$/d$_0$~$\approx$\,-10~\%). Similar
deviations have also been observed with Ru\,(0001) [30,31]. The
inclusion of the effects of zero-point vibration and thermal expansion
[32] in the calculations further reduces the difference between theory
and experiment: at T = 300K d$_{12}$/d$_{0}$ = $-$12.3~\% which agrees
with the experiments within the error bars. The calculated work
function of 4.88~eV is somewhat too low compared to the experiment,
5.10~eV [33], again as also observed in DFT-GGA calculations of other
metals [29].

  To study the adsorption of oxygen, we have considered several
coverages and overlayer arrangements. They are compiled in Table~3,
together with the adsorption energies per O atom and work function
changes with respect to the clean surface. In addition to the
experimentally observed surface structures, we also studied some
hypothetical structures in order to gain additional information about
the adsorbate-adsorbate interaction. From Table~3, we see that the
adsorption energy decreases upon oxygen adsorption, while the work
function increases. Even the (1$\times$1)-2O phase is found to be
stable. This is no surprise, as the oxygen atoms are bound on
threefold hollow sites as on the Ru\,(0001) surface, and the density
of adsorbates is close to the one in the Ru\,(0001)-(1$\times$1)-O
structure, which was also found to be stable [15,31]. Interesting is
that the energetically lowest (1$\times$1)-2O structure on
Ru\,(10$\bar{1}$0) contains both atoms adsorbed on the hcp and fcc
sites, whereas in the (1$\times$1)-1O structure the hcp site is
clearly preferred. The occupation of fcc and hcp sites in the
(1$\times$1)-2O phase maximizes the separation between the O atoms
within the troughs and therefore minimizes the electrostatic O-O
repulsion. Yet, the LEED experiments did not give evidence for the
presence of a (1$\times$1)-2O structure. Therefore, one might
conjecture that the excess oxygen atoms partly penetrate into the
subsurface region at a sample temperature of 550~K leading to the
diffuse and streaky (1$\times$2) LEED pattern.

  Comparing the nearest-neighbor distance and the adsorption energy of
the (2$\times$1)pg-2O with the (1$\times$1)-1O, we find the energy
gain by forming zigzag chains to be 60~meV, which is consistent with a
reduction of the O-O repulsion. The binding energy per adsorbate
increases by 220~meV per atom, when the coverage is reduced from 1 to
1/2. However, the c(2$\times$4)-2O structure, where the zigzag and
zagzig chains alternate, is preferred over the (2$\times$2)-2O
exhibiting zigzag chains only. This might be related to a more
favorable relaxation of the outermost Ru atoms in the trenches.

  Next, we are focusing on the oxygen-Ru distances as a function of
coverage. The O-Ru distances in the c(2$\times$4) (d[O-Ru(1)] =
2.09~\AA\ and d[O-Ru(2)] = 2.10~\AA) and the (2$\times$1)pg structures
(d[O-Ru(1)] = 2.11~\AA\ and d[O-Ru(2)] = 2.11~\AA) are larger than in
the experiment (c(2$\times$4)-2O: d[O-Ru(1)] = 2.08~\AA/d[O-Ru(2)] =
2.03~\AA; (2$\times$1)pg-2O: (d[O-Ru(1)] = 2.03~\AA\ and d[O-Ru(2)] =
2.03~\AA)), which might be due to the DFT-GGA scheme. In disagreement
with the experiments, the bond lengths vary only slightly, when going
from the c(2$\times$4)-2O to the (2$\times$1)-2O structure; the LEED
analyses show a notable and clear decrease of the O-Ru(1) bond length.
Specific values for the atomic coordinates as determined by DFT-GGA
calculations are indicated in Figs. 4 and 6. However, going to even
higher O coverages, also DFT-GGA calculations indicate a decreasing
O-Ru distance. DFT-GGA calculations of the system O/Ru\,(0001)
[31,~34] show that the O-Ru bond strength decreases monotonically with
increasing O coverage.

  One should note that the Ru-O binding energy is the results of both
the direct O-Ru interaction and the O-O repulsion. It is well possible
that the direct O-Ru interaction increases with coverage (consistent
with a shorter bond length) but the adsorption energy decreases,
because of the electrostatic O-O repulsion. Unfortunately both
contributions cannot be separated in the DFT calculations. A more
detailed discussion of this effect as found with alkali metal
adsorption on Al\,(111) may be found in Ref. 21.

  Using the same pseudo potentials as for O/Ru\,(10$\bar{1}$0), the
binding energies of oxygen on Ru\,(0001) turned out to be 3.08~eV and
2.29~eV and the Ru-O bond lengths are 2.07~\AA\ and 2.04~\AA\ in the
(2$\times$2)-O and the (1$\times$1)-O phase, respectively. The
slightly larger Ru-O binding energy in the Ru\,(0001)-(2$\times$2)-O
phase compared to Ru\,(10$\bar{1}$0)-c(2$\times$4)-O is due to the
larger O-O separation.

  In addition to the above DFT-GGA calculations, we analyzed the
binding of oxygen on Ru\,(10$\bar{1}$0) by using a simple
tight-binding model [35], which we recently applied to N/Ru\,(0001)
[22]. The qualitative result is the same (cf. Fig.~9) {\it i.\,e.},
the bond splits the mixed adsorbate-substrate orbitals - mainly
O$_{2p}$ and Ru$_{4d}$ - into a bonding level below and an
anti-bonding level above the Fermi level, emptying some Ru4d orbitals
just below the Fermi energy. The position of the bonding orbital
shifts from E$_F - 5.5$~eV [c(2$\times$4)-2O] to E$_F - 6.5$~eV
[(1$\times$1)-2O] with increasing coverage; the position of the
bonding orbital does not shift, when going from the c(2$\times$4)-2O
to the (2$\times$1)pg-2O. The anti-bonding peak is at E$_F + 1.5$~eV
in the c(2$\times$4)-2O structure and shifts slightly to higher
energies when increasing the coverage. Both shifts are in agreement
with the smaller O-Ru distance upon increasing the O coverage (taken
from Table~3), if one neglects the effective repulsion between the
oxygen atoms.

\section{Discussion}

  The interaction of oxygen with the substrate and neighboring oxygen
atoms can be classified in three categories, which will be described
in the sequence of their importance.

1. Oxygen binds to the Ru substrate via two atoms in the topmost Ru
layer and one in the second layer. Such an adsorption site may also be
the adsorption site in the limit of zero coverage. Obviously, oxygen
prefers to form bonds to the low-coordinated Ru atoms of the topmost
layer. According to Tersoff and Falicov [36], this adsorption behavior
might be related to the local d-band of these Ru atoms. Since the
coordination number of these atoms is smaller than that of the other
Ru atoms, the local d-band is narrower (in the tight-binding model the
band width is proportional to the square root of the coordination
number) and since the Fermi level is located about in the middle of
the band, the local density of states becomes larger. Accordingly, Ru
atoms in the second layer (with the higher coordination) should have a
smaller density of states than Ru atoms in the outermost layer. Using
now simple effective medium theory (EMT) arguments, which have been
put forward by N{\o}rskov and coworkers [1a,37], the bonding between
oxygen and Ru depends critically on the local charge density offered
by the substrate: The more local charge density is available, the
longer the O-Ru bond length (and the smaller is the Pauli repulsion).
In fact, this situation is encountered with the c(2$\times$4)-2O
overlayer. The bond length of O to Ru(1) with 2.08~\AA\ is markedly
longer than that between O and Ru(2) (2.03~\AA). In addition, the
O-induced reconstructions are much less pronounced on
Ru\,(10$\bar{1}$0) than on Ru\,(0001) [38]. This finding is consistent
with the notion that oxygen atoms on Ru\,(10$\bar{1}$0) can readily
reach the optimum charge density without shifting the Ru atoms that
much. The opposite is the case for Ru\,(0001), since all Ru atoms are
high-coordinated and, therefore, the local charge density is low.
Consequently, the O-Ru bond length of Ru\,(0001)-(2$\times$2)-O
(2.03~\AA\ [38]) is quite short and comparable to the O-Ru(2) value on
Ru\,(10$\bar{1}$0). Yet, this interpretation within the framework of
local charge density is not conclusive, as the present DFT
calculations (which are certainly superior over EMT calculations in
terms of accuracy) cannot reconcile this behavior for reasons which
are unknown: There is almost no variation in the O-Ru bond length with
the number of O atoms coordinated to a Ru atom. DFT calculations do
show that the hcp adsorption site of oxygen is favored by more than
0.2~eV over the fcc site (cf. Table~3, [1$\times$1)1O-fcc and
(1$\times$1)1O-hcp].

2.      The next important contribution to the Ru-O bond establishes
the formation of zigzag chains along the troughs. These zigzag
arrangements were also observed for hydrogen on fcc\,(110) surfaces,
such as Ni\,(110) [39] and oxygen on Rh\,(110) [40], and it therefore
seems a quite common structural motif in surface science. LEED and
HREELS measurements indicate that even at lower O coverages these
chains are preferentially formed. This observation is also supported
by the DFT calculations which indicated that the formation of zigzag
chains is by 60~meV more favorable than single O atoms. This means
that the interaction between the oxygen atoms along the zigzag chains
is attractive. Albeit this attractive interaction, oxygen does not
like to share attached Ru atoms with other oxygen atoms. This
attribute is manifested by the empty trough between the zigzag chains
observed in the c(2$\times$4)-2O system. If, however, the O coverage
is increased, the empty troughs are filled up with the consequence
that oxygen atoms share common Ru atoms in the topmost layer.
Accordingly, the local charge density offered by Ru(1) atoms for each
O atoms is smaller and, again using the simple effective medium theory
of bonding, this tells us that now the oxygen atoms have to come
closer to the surface to experience a similar local charge density as
in the c(2$\times$4)-2O phase. In fact, this aspect has been
identified in the LEED analysis, the bond lengths of O to Ru(1) change
from 2.08~\AA to 2.01~\AA, when going from the c(2$\times$4)-2O to the
(2$\times$1)-2O phase. The bonding between the oxygen atom and the
coordinated Ru(2) atom is not affected by the presence of more oxygen
on the surface, as evidenced by the unchanged bond length of 2.03~\AA.
These findings cannot be easily reconciled with simple arguments
adopted from coordination chemistry, since there the bond length is
directly related to the bond strength: The weaker the bonding, the
longer the respective bond length. The present DFT calculations show,
however, that the binding energy of oxygen is reduced by about
220~meV, when the oxygen coverage is increased so that two oxygen
atoms have to share one Ru atom in the topmost layer. Accordingly, the
O-Ru(1) and O-Ru(2) bond lengths should increase which, however, is
not observed experimentally. One should recall that the Ru-O binding
energy is composed of the direct O-Ru interaction and the O-O
repulsion. It is conceivable that the direct O-Ru interaction
increases with coverage (consistent with a shorter bond length) but
the adsorption energy decreases, because of the electrostatic O-O
repulsion. Since we are not able to entangle these two contributions
in our DFT calculations this argument remains elusive.

  Unfortunately, the present DFT calculations are not able to
elaborate on this point, as the optimum bond length turned out not to
change with O coverage when the c(2x4)-2O phase transforms into the
(2x1)pmg-2O phase. It could be that such an effect is beyond the
capability of state-of-the-art DFT-GGA calculations. As possible
sources for this discrepancy we mention just the exchange correlation
term, the frozen-core approximation, the linearization of the
core-valence exchange correlation and the transferability of the
pseudopotentials in our DFT calculations. Although the variation of
the Ru(1)-O bond length with O-coverage (as determined by LEED) is
similar to the quoted error bars for the Ru(1)-O bond lengths, simple
probability arguments, {\it i.\,e.} taking the product probability,
tell that the found variation is statistically significant. A simple
calculation shows that the probability to find the Ru(2)-O bond length
to increase is only 20~\%.

  An alternative explanation for the observed shortening of the
Ru(2)-O bond length with O-coverage could be that with increasing O
coverage the ionicity of O decreases. Consequently, the Pauling radius
of oxygen would shrink consistent with a smaller Ru-O bond length.
This explanation fails however to explain the observed change of work
function $\Delta\phi$. Using ultraviolet photoelectron spectroscopy
[41] it was shown that $\Delta\phi$ increases from 0.49~eV [c(2x4)-2O]
to 1.12~eV [(2x1)pg-2O] with respect to the clean surface. This
super-linear increase of the oxygen induced dipole moment conflicts
with a reduced degree of ionicity in combination with a smaller Ru-O
bond length, {\it i.\,e.} smaller dipole length. It is worth
mentioning that this super-linear change of the work function induced
by oxygen is nicely reproduced by the present DFT calculations.
Drastic changes in the oxygen induced dipole moment with coverage
signifies the importance of the electrostatic repulsion between the O
atoms. It might be that this adsorbate-adsorbate interaction give a
clue to the experimentally observed shortening of Ru-O bond length
(cf. the discussion in Ref. [21]).

  A quite similar effect has recently been observed for oxygen on
Ru\,(0001) [15]. While the O-Ru bond length in both the (2$\times$2)
and the (2$\times$1) phase was 2.03~\AA\ [38], this bond length
shortens to 2.00~\AA\ for the (1$\times$1)-O structure, although the
binding energy decreases by more than 0.8~eV, as evidenced by DFT
calculations. Obviously, local coordination chemistry considerations,
which would predict that the stronger the bond, the shorter the bond
length, are inadequate to account for the bonding of oxygen atoms at
the Ru surfaces. Interestingly, on Ru\,(0001) the $\nu$(Ru-O) stretch
mode energy increases with oxygen coverage, as found by recent HREELS
study on Ru\,(0001) [28]. Hence, the Ru-O potential well obviously
becomes steeper, when the Ru-O distance decreases.

  On Ru\,(10$\bar{1}$0) the Ru-O stretch mode changes its energy only
slightly from 67 to 64~meV with coverage, {\it i.\,e.}, the steepness
of the potential normal to the surface does not change very much. The
frequency is very close to that one for the low-coverage oxygen phase
on Ru\,(0001) (64 meV at $\theta_{\rm O}$~= 0.25). Only when the
lateral nearest-neighbor O-O distance on Ru\,(0001) changes from
5.72~\AA\ ($\theta_{\rm O}$~= 0.25) to 2.71~\AA, the $\nu$(Ru-O) mode
energy changes to 71~meV ($\theta_{\rm O}$~= 0.50) and 81~meV
($\theta_{\rm O}$~= 1.00) [28]. For Ru\,(10$\bar{1}$0) the
nearest-neighbor distances remain quite large in both phases:
3.30~\AA\ in the c(2$\times$4)-2O and 3.38~\AA\ in the
(2$\times$1)-2O. Therefore, the $\nu$(Ru-O) mode energy is nearly
constant. The small change from 67~meV to 64~meV goes along with an
increase of the nearest-neighbor distance, similar to the effect
observed with Ru\,(0001).

  Finally, we like to note that for oxygen, in contrast to N and H on
Ru\,(10$\bar{1}$0), basically only the mode perpendicular to the
surface is visible. For N [42] and H [43] at least one translational
mode was observed in addition. We conclude that for oxygen the bond is
more symmetrical, with respect to the surface normal, than for N and
H. This conclusion is also supported by recent LEED analyses of H on
Re(10$\bar{1}$0) and Ru\,(10$\bar{1}$0), where the short-bridge site
has been identified [26].

  From the HREELS results - that the O against Ru vibration frequency
shifts from 67~meV to 64~meV, when going from the c(2$\times$4)-2O to
(2$\times$1)pg-2O - one can learn that the potential perpendicular to
the surface varies more softly in the (2$\times$1)pg-2O than in the
c(2$\times$4)-2O phase. This might be attributed to the smaller bond
strength of oxygen to the Ru surface in the (2$\times$1)pg-2O than in
c(2$\times$4)-2O, as indicated by the present DFT calculation.

3.      The smallest interaction energy is that which determines the
correlation between the zigzag chains in the c(2$\times$4)-2O phase.
This interaction can easily be overcome by increasing the surface
temperature above 550~K or simply adding more oxygen to the surface.
DFT calculations indicate that this energy contribution is with 60~meV
indeed quite small [cf. Table~3: c(2$\times$4)-2O and
(2$\times$2)-2O].

\section{Summary}

  The c(2$\times$4)-2O and the (2$\times$1)pg-2O phases on
Ru\,(10$\bar{1}$0) were characterized by using quantitative LEED,
DFT-GGA calculations and HREELS. We have shown that in both phases
oxygen atoms occupy the threefold coordinated hcp site along the
densely packed rows on an otherwise unreconstructed surface, {\it
i.\,e.}, the O atoms are attached to two atoms in the first Ru layer
Ru(1) and to one Ru atom in the second layer Ru(2). With LEED we found
that in the low-coverage c(2$\times$4)-O phase the bond lengths of O
to Ru(1) and Ru(2) are 2.08~\AA\ and 2.03~\AA, respectively, while
corresponding bond lengths in the high-coverage (2$\times$1)-2O phase
are 2.01~\AA\ and 2.04~\AA. The shortening of the Ru(1)-O bond length
with O coverage may be a consequence of the competition for electron
charge density, although this aspect could not be reconciled by our
DFT calculations. The presence of empty troughs in the c(2x4)-2O phase
indicates that oxygen atoms do not like to share Ru atoms in the
topmost layer with other O atoms. DFT calculations show that the
energy per O atom in both phases differs by 220~meV. The energy gain
of 60~meV drives the O atoms to build zigzag chains along the troughs
instead of a dispersed O-phase. A similar energy contribution is
gained when forming alternating zigzag and zagzig chains instead of
zigzag chains only. HREEL spectra reveal a loss at 67~meV
[c(2$\times$4)-2O] and 64~meV [(2$\times$1)pg-2O], which is assigned
to the $\nu$(Ru-O) stretch mode. The small variation in frequency with
O coverage [compared to a much bigger shift for the oxygen phases on
Ru\,(0001)] reflects the similar lateral arrangements of oxygen in
both phases on Ru\,(10$\bar{1}$0), {\it i.\,e.} the formation of
zigzag chains.

We thank B. Hammer and R. Stumpf for valuable discussions. V.~D.
acknowledges partial financial support by the A.~Della Riccia
Foundation.

\section{References}
[1] see {\it e.\,g.} a) J. K. N{\o}rskov, Rep. Prog. Phys. {\bf 53},
1253 (1990); b) M.J. Stott and E. Zaremba, Phys. Rev. B {\bf 22}, 1564
(1980); c) M.W. Finnis and J.E. Sinclair, Philos. Mag. A {\bf 50}, 45
(1984), d) F. Ducastelle and F. Cyrot-Lackmann, J. Phys. Chem. Solids
{\bf 32}, 285 (1971); e) F. Ercolessi, E. Tosatti, and M. Parrinello,
Phys. Rev. Lett. {\bf 57}, 719 (1986); f) I.J. Robertson, M.C. Payne,
and V. Heine, Phys. Rev. Lett. {\bf 70}, 1944 (1993).

[2] a) B. Narloch, G. Held, and D. Menzel, Surf. Sci. {\bf 317}, 131
(1994); b) H. Over, H. Bludau, R. Kose, and G. Ertl, Phys. Rev. B {\bf
51}, 4661 (1995).

[3] a) H. Over, H. Bludau, M. Skottke-Klein, G. Ertl, W. Moritz,
and C.T. Campbell, Phys. Rev. B {\bf 45}, 8638 (1992); b) M. Gierer, H.
Bludau, T. Hertel, H. Over, W. Moritz, and G. Ertl, Surf. Sci. {\bf 279},
L170 (1992).

[4] C. Stampfl and M. Scheffler, Surf. Rev. Lett. {\bf 2}, 317 (1995).

[5] R. Koch, B. Burg, K.-H. Rieder, and E. Schwarz, Mod. Phys. Lett. B
{\bf 8}, 571 (1994).

[6] E. Schwarz, K.H. Schwarz, C. Gonser-Buntrock, M. Neuber, and
K.~Christmann, Vacuum {\bf 41}, 180 (1990).

[7] M. Gierer, H. Over, P. Rech, E. Schwarz, and K. Christmann, Surf.
Sci. {\bf 370}, L201 (1997).

[8] T. W. Orent and R. S. Hansen, Surf. Sci. {\bf 67}, 325 (1977).

[9] R. Koch, E. Schwarz, K. Schmidt, B. Burg, K. Christmann, and K.-H.
Rieder, Phys. Rev. Lett. {\bf 71}, 1047 (1993).

[10] K. M\"uller, E. Lang, L. Hammer, W. Grimm, P. Heilmann, and K.
Heinz, in: Determination of Surface Structure by LEED, edited by P.M.
Marcus and F. Jona (Plenum, New York, 1984).

[11] W. Moritz, J. Phys. C. {\bf 17}, 353 (1983).

[12] a) G. Kleinle, W. Moritz, and G. Ertl, Surf. Sci. {\bf 238}, 119
(1990); b) H. Over, U.~Ketterl, W. Moritz, and G. Ertl, Phys. Rev. B
{\bf 46}, 15438 (1992); c) M.~Gierer, H. Over, and W. Moritz,
unpublished.

[13] J. B. Pendry, J. Phys. C {\bf 13}, 937 (1980).

[14] G. Kleinle, W. Moritz, D. L. Adams, and G. Ertl, Surf. Sci. 
{\bf 219}, L637 (1989).

[15] C. Stampfl, S. Schwegmann, H. Over, M. Scheffler, and G. Ertl,
Phys. Rev. Lett. {\bf 77}, 3371 (1996).

[16] J. P. Perdew, J. A. Chevary, S. H. Vosko, K. A. Jackson, M. R.
Pederson, D. J. Singh and C. Fiolhais, Phys. Rev. B {\bf 46}, 6671
(1992).

[17] N. Troullier and J. L. Martins, Phys. Rev. B {\bf 43}, 1991
(1993).

[18] The binding energy and bond length of O$_2$ in the gas phase are
well converged while the vibrational frequency is not. The Ru-O bonds
at the surface are also well converged and the O-O interaction on the
surface is partially screened by the substrate and not important due
to the large O-O separation.

[19] Note that in the DFT calculation of Ref. [31] the same pseudo
potential for Ru, but a slightly different pseudo potential for oxygen
was taken (r$_c$$^{l=0,1}$ = 1.35 bohr).

[20] H. J. Monkhorst and J. D. Pack, Phys. Rev. B {\bf 13}, 5188
(1976).

[21] J. Neugebauer and M. Scheffler, Phys. Rev. B {\bf 46}, 16067
(1992).

[22] S. Schwegmann, A. P. Seitsonen, H. Dietrich, H. Bludau, H. Over,
K. Jacobi, and G. Ertl, Chem. Phys. Lett. {\bf 264}, 680 (1997).

[23] S. Schwegmann, V. De Renzi, H. Bludau, H. Over, and G. Ertl,
unpublished.

[24] H. L. Davis and D. M. Zehnder, J. Vac. Sci. Technol. {\bf 17},
190 (1980).

[25] H. Over, G. Kleinle, W. Moritz, G. Ertl, K.-H. Ernst, H.
Wohlgemuth, K. Christmann, and E. Schwarz, Surf. Sci. {\bf 254}, L469
(1991).

[26] R. D\"oll, L. Hammer, K. Heinz, K. Bed\"urftig, U. Muschiol, K.
Christmann, H. Bludau and H. Over, in preparation.

[27] W. J. Mitchell and W. H. Weinberg, J. Chem. Phys. {\bf 104}, 9127
(1996).

[28] P. He and K. Jacobi, Phys. Rev. B {\bf 55}, 4751 (1997).

[29] M. Fuchs and M. Scheffler, to be published.

[30] G. Michalk, W. Moritz, H. Pfn\"ur and D. Menzel, Surf. Sci. {\bf
129}, 92 (1983).

[31] C. Stampfl and M. Scheffler, Phys. Rev. B {\bf 54}, 2868 (1996).

[32] a) J.-H. Cho and M. Scheffler, Phys. Rev. Lett. {\bf 78} (1997)
1299; b) S. Narasimhan and M. Scheffler, to be published.

[33] Landolt B\"ornstein, Group III Vol. 24, Subvolume B (Springer,
Berlin-Heidelberg, 1994) p. 277.

[34] A. P. Seitsonen, unpublished calculations.

[35] R. Hoffmann, Rev. Mod. Phys. {\bf 60}, 601 (1988).

[36] J. Tersoff and L. M. Facilov, Phys. Rev. B {\bf 24}, 754 (1980).

[37] a) K. W. Jacobsen, J. K. N{\o}rskov, and M. J. Puska, Phys. Rev.
B {\bf 35}, 7423 (1987); b) B. Hammer and J. K. N{\o}rskov, in: Theory
of Adsorption and Surface Reactions, NATO ASI, Eds. R. Lambert and G.
Pacchioni (Kluwer, Amsterdam, 1997).

[38] a) H. Pfn\"ur, G. Held, M. Lindroos, and D. Menzel, Surf. Sci.
{\bf 220}, 43 (1990); b) M. Lindroos, H. Pfn\"ur, G. Held, and D.
Menzel, Surf. Sci. {\bf 222}, 451 (1989).

[39] V. Penka, K. Christmann, and G. Ertl, Surf. Sci. {\bf 136}, 307
(1984).

[40] M. Gierer, H. Over, G. Ertl, H. Wohlgemuth, E. Schwarz, and
K.~Christmann, Surf. Sci. Lett. {\bf 279}, L73 (1993).

[41] P. Rech, PhD thesis, FU Berlin 1996, Germany.

[42] H. Dietrich, K. Jacobi, and G. Ertl, J. Chem. Phys. (accepted).

[43] M. Gryters and K. Jacobi, J. Electron. Spectrosc. Relat. Phenom.
{\bf 64/65}, 591 (1993).

\clearpage

\section{Figure Captions}

FIG. 1. LEED patterns of the a) clean Ru\,(10$\bar{1}$0), b)
Ru\,(10$\bar{1}$0)-c(2$\times$4)-2O, and
c)~Ru\,(10$\bar{1}$0)-(2$\times$1)pg-2O.

FIG. 2. Possible structure models for the
Ru\,(10$\bar{1}$0)-c(2$\times$4)-2O, which were tested by LEED
calculations.

FIG. 3. Possible structure models for the
Ru\,(10$\bar{1}$0)-(2$\times$1)pg-2O, which were tested by LEED
calculations.

FIG. 4. The atomic coordinates for the best-fit model of
c(2$\times$4)-2O as obtained by quantitative LEED and DFT-GGA
calculations. Oxygen atoms reside in hcp-like adsorption sites forming
zigzag chains along the troughs. Zigzag and zagzig chains are
separated by empty troughs so that no oxygen atom has to share a Ru
atom in the topmost layer with neighboring O atoms. The oxygen bond
lengths to first-layer and second-layer Ru atoms are 2.08~\AA\ and
2.03~\AA, respectively.

FIG. 5. Comparison of experimental and theoretical LEED IV data for
the best-fit model of the c(2$\times$4)-2O phase on Ru\,(10$\bar{1}$0)
(cf. Fig.~4). The overall R$_P$-factor is 0.26.

FIG. 6. The atomic coordinates for the best-fit model of
(2$\times$1)pg-2O, as determined by LEED and DFT. Oxygen atoms reside
in hcp-like adsorption sites forming zigzag chains along the troughs.
The oxygen bond lengths to first-layer and second-layer Ru atoms are
2.01~\AA\ and 2.04~\AA, respectively.

FIG. 7. Comparison of experimental and theoretical LEED IV data for
the best-fit model of the (2$\times$1)pg-2O phase on
Ru\,(10$\bar{1}$0) (cf. Fig.~6). The overall R$_P$-factor is 0.25.

FIG. 8. HREEL spectra for a series of oxygen exposures on
Ru\,(10$\bar{1}$0) at room temperature. The exposures and the
monitored LEED patterns are indicated in the figure. All spectra are
recorded in specular geometry with primary energy of 2.5~eV. The
scaling factor in the loss region is 100.

FIG. 9. The difference in density of states, n(Ru+O)~-~n(Ru), where
n(Ru+O) is the density of states of the adsorbate system
Ru(10$\bar{1}$0)-(2$\times$1)pmg-O (doted-dashed line),
Ru(10$\bar{1}$0)-(1$\times$1)-(fcc+hcp)2O (solid line),
Ru(10$\bar{1}$0)-c(2$\times$4)-2O (dotted line) and n(Ru) that of the
clean Ru surface. The one-electron eigenvalues are broadened by
0.6~eV.

\section{Tables}

\vspace{5mm}

Table 1: Optimum Pendry-r-factors obtained for different models of the
Ru(10$\bar{1}$0)-2O-c(2$\times$4). The total energy range is 4525~eV
(2445~eV fractional-order, 2080~eV integer-order beams).
\begin{center}
\begin{tabular}{|l||c|c|c|}\tableline
                 &   \multicolumn{3}{c|}{R$_p$} \\
adsorption sites &   total &          integer & fractional \\\tableline
fcc A            &   0.65  &          0.30 &   0.94 \\
fcc B            &   0.66  &          0.30 &   0.94 \\
fcc C            &   0.66  &          0.29 &   0.96 \\
hcp A            & \underline{\bf 0.26} & {\bf 0.20} & {\bf 0.33} \\
hcp B            &   0.57  &          0.30 &   0.83 \\
hcp C            &   0.50  &          0.24 &   0.76 \\
long bridge      &   0.74  &          0.39 &   1.02 \\
short bridge     &   0.68  &          0.42 &   0.91 \\\tableline
\end{tabular}
\end{center}

\vspace{5mm}

Table 2: Optimum Pendry-r-factors obtained for different models of the
Ru(10$\bar{1}$0)-2O-(2$\times$1)p2mg. The total energy range is
3621~eV (1464~eV fractional-order, 2157~eV integer-order beams).
\begin{center}
\begin{tabular}{|l||c|c|c|}\tableline
                 &   \multicolumn{3}{c|}{R$_p$} \\
adsorption sites & total & integer & fractional \\\tableline\tableline
fcc              & 0.71  & 0.55    & 0.95 \\
hcp              & \underline{\bf 0.25} & {\bf 0.23} & {\bf 0.29} \\
short bridge     & 0.69  & 0.56    & 0.87 \\\tableline
\end{tabular}
\end{center}

\end{multicols}

\vspace{5mm}

Table 3: Oxygen adsorption energy, work function change and O-Ru
distances in different overlayer structures. The adsorption energy is
calculated with respect to a free oxygen molecule. Ru(1) denotes the
outermost-layer Ru atoms and Ru(2) the second-layer atoms. In the
structure (1$\times$1)-2O (hcp-fcc) we give the average of the
distances from the atom at the hcp/fcc site to the Ru atoms.
\begin{center}
\begin{tabular}{|l||c|c|c|c|c|}\tableline
structure & coverage & E$_{\rm ad}$ (eV) & $\Delta\phi$ (eV) &
     d[O-Ru(1)] (\AA) & d[O-Ru(2)] (\AA) \\ \tableline\tableline
c(2$\times$4)-2O        & 0.50 & 2.81 & 0.69 & 2.09 & 2.10 \\
(2$\times$2)-2O         & 0.50 & 2.75 & 0.65 & 2.08 & 2.11 \\
(2$\times$1)-1O         & 0.50 & 2.75 & 0.69 & 2.07 & 2.14 \\
(2$\times$1)pmg-2O      & 1.00 & 2.59 & 1.61 & 2.11 & 2.11 \\
(1$\times$1)-1O hcp     & 1.00 & 2.53 & 1.16 & 2.07 & 2.09 \\
(1$\times$1)-1O fcc     & 1.00 & 2.30 & 0.74 & 2.04 & 2.11 \\
(1$\times$1)-2O hcp     & 2.00 & 1.37 & 1.33 & 2.01 & 2.10 \\
(1$\times$1)-2O hcp-fcc & 2.00 & 1.47 & 1.91 & 1.98 & 2.08 \\ \tableline
\end{tabular}
\end{center}

\end{document}